\documentclass[sigchi,nonacm]{acmart}
\AtBeginDocument{%
  }

\setcopyright{acmlicensed}
\copyrightyear{2025}
\acmYear{2025}
\acmDOI{XXXXXXX.XXXXXXX}
\acmConference[Conference acronym 'XX]{Make sure to enter the correct
  conference title from your rights confirmation email}{June 03--05,
  2025}{Woodstock, NY}

\begin{document}

\title[Investigating Environments’ and Avatars’ Effects on Thermal Perception in VR]{Investigating Environments’ and Avatars’ Effects on Thermal Perception in Virtual Reality to
Reduce Energy Consumption}


\author{Martin Kocur}
\orcid{0000-0003-3077-6267}
\affiliation{%
  \institution{University of Applied Sciences Upper Austria}
  \city{Hagenberg}
  \country{Austria}
}
\email{martin.kocur@fh-hagenberg.at}

\author{Niels Henze}
\orcid{0000-0002-5702-6290}
\affiliation{%
	\institution{University of Regensburg}
	\city{Regensburg}
	\country{Germany}
}
\email{niels.henze@ur.de}

\renewcommand{\shortauthors}{Kocur and Henze}

\begin{abstract}
Understanding thermal regulation and subjective perception of temperature is crucial for improving thermal comfort and human energy consumption in times of global warming. Previous work shows that an environment’s color temperature affects the experienced temperature. As virtual reality (VR) enables visual immersion, recent work suggests that a VR scene's color temperature also affects experienced temperature. In addition, virtual avatars representing thermal cues influence users’ thermal perception and even the body temperature. As immersive technology becomes increasingly prevalent in daily life, leveraging thermal cues to enhance thermal comfort—--without relying on actual thermal energy—--presents a promising opportunity. Understanding these effects is crucial for optimizing virtual experiences and promoting sustainable energy practices. Therefore, we propose three controlled experiments to learn more about thermal effects caused by virtual worlds and avatars.

\begin{figure*}
    \centering
    \includegraphics[width=0.49\textwidth]{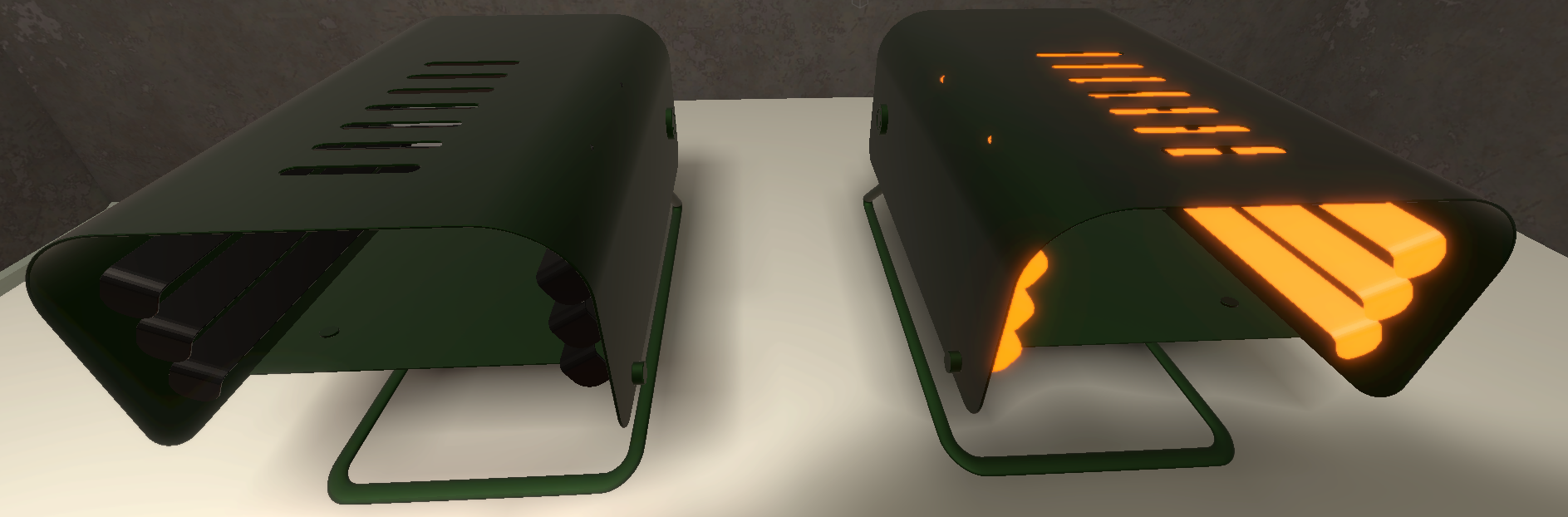}
    \includegraphics[width=0.4\textwidth,trim = 0cm 2cm 0cm 2.4cm,clip]{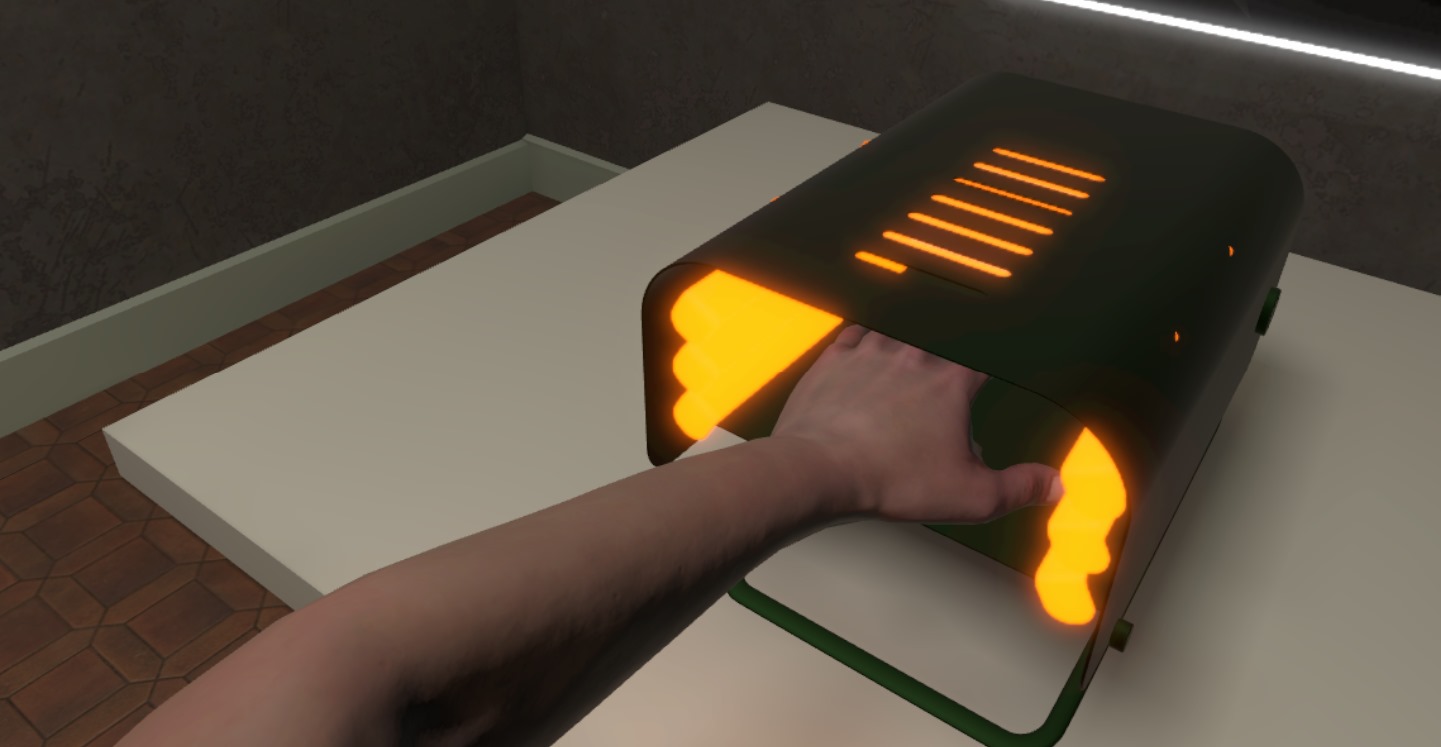}
  \caption{We aim to investigate whether thermal cues, such as virtual heaters, influence thermal perception and skin temperature during a VR experience. We hypothesize that activating the virtual heater will lead to an increase in skin temperature compared to when it is turned off.}
  \label{fig:heater}
\end{figure*}
\end{abstract}

\ccsdesc[500]{Human-centered computing~Virtual reality}
\ccsdesc[300]{Applied computing~Computer games}

\keywords{virtual reality, mixed reality, climate change, proteus effect, virtual embodiment, thermal comfort}
\begin{teaserfigure}
  \includegraphics[width=\textwidth]{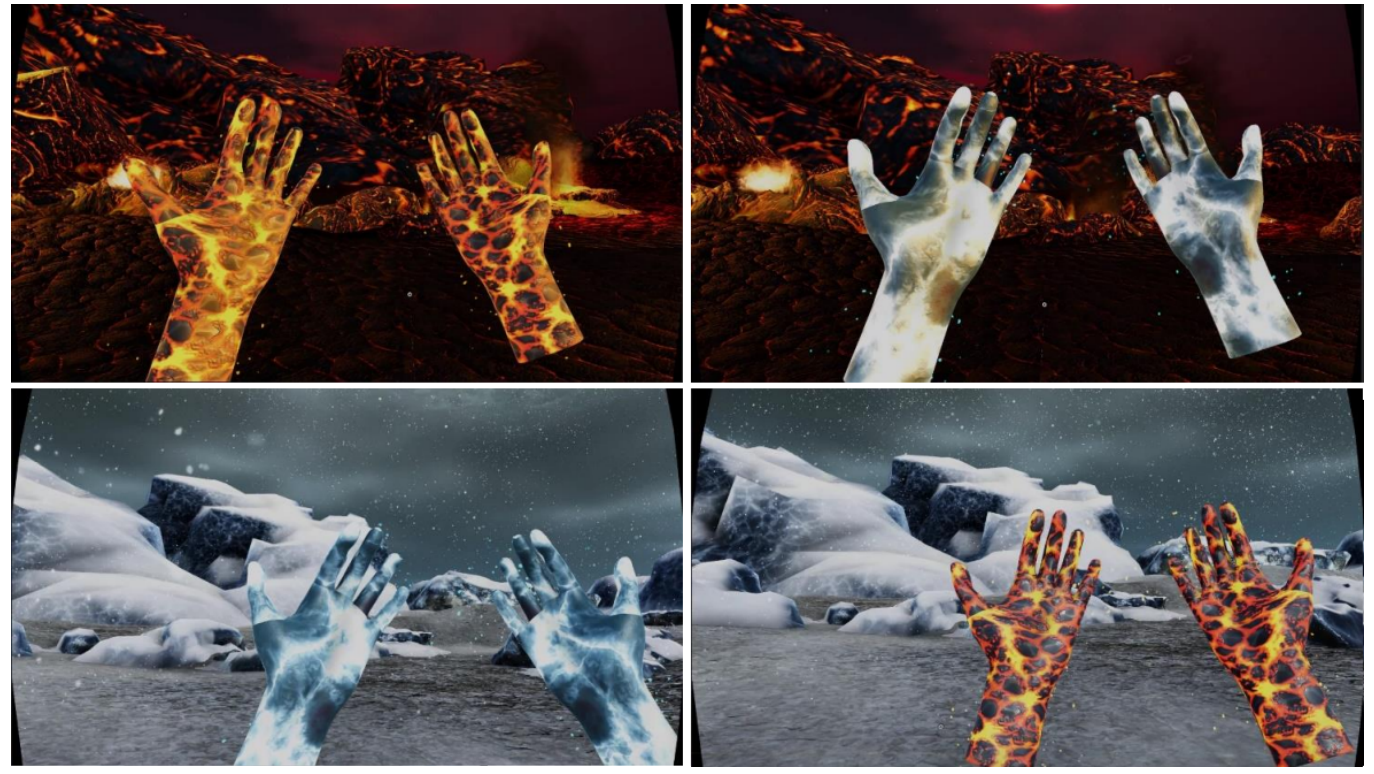}
  \caption{Previous Work: Users embodying fire or ice avatar hands while being immersed in a fire or ice world~\cite{KocurHenze23}. Participants had a reduced skin temperature while embodying fire hands compared to ice hands. The virtual environment had no effects on skin temperature, however, systematically influenced thermal perception: Participants felt warmed in the fire world compared to the ice world.}
  \label{fig:teaser}
\end{teaserfigure}

\maketitle
\section{Introduction and Background}

Climate change is one of the greatest challenges of the 21st century and is caused by the increased concentration of greenhouse gases in the atmosphere, primarily due to human activities~\cite{Trenberth18}. While climate change has severe consequences on the environment such as long-term changes in temperature, precipitation patterns, and sea levels~\cite{Thuiller2011}, it also has a negative impact on people's health, especially due to the increase in heatwaves and extreme weather events caused by global warming~\cite{ClimateChangeHuman}. Therefore, there is an urgent need to fight climate change and contribute to its mitigation~\cite{UNESCO,Lemons2011}.

Experts agree that fighting climate change requires a collective effort from all disciplines~\cite{Molthan-Hill2022}. Traditionally, climatologists and environmental scientists focus on studying the impact of climate change and guide strategies of policymakers. However, other research disciplines are also engaged to make valuable contributions, e.g., psychology, biology, business, physics, and computer science ~\cite{Molthan-Hill2022}. In recent years, the field of human-computer interaction (HCI) started to explore how digital technology can raise awareness for environmental issues and promote a sustainable lifestyle to combat climate change~\cite{Knowles14,Mencarini23,Knowles18}. 

 
The most significant factor contributing to man-made climate change is a high energy consumption~\cite{AKHMAT2014123}. Heating or cooling buildings to satisfy the human need for comfortable ambient temperatures requires enormous energy consumption and has a direct environmental impact~\cite{BALARAS2005429}. While the debate about more efficient solutions to reach a specific ambient temperature has gained momentum, e.g., district heating~\cite{ZHANG2021116605} or heat pumps~\cite{HeatPump18}, such alternatives still require energy and can have a carbon footprint~\cite{MIKIELEWICZ2022123280}. As physically changing the temperature always requires a certain amount of energy according to the law of thermodynamics~\cite{ALEFELD1987331,URGEVORSATZ201585}, we need to find ways to reduce heating or cooling while still maximizing thermal comfort and ensuring that temperatures are perceived as pleasant. 

Previous work showed that one's thermal comfort is malleable and does not only depend on the actual temperature~\cite{kim2021influencing}. The hue-heat hypothesis, for example, suggests that color temperature affects temperature perception and thermal comfort~\cite{mogensen26}. Recent work found that lighting's color temperature influences thermal comfort~\cite{huebner2016saving, alfano2019thermal, bellia2021interaction} and light is even capable of evoking thermophysiological responses~\cite{te2016influence}. Overall, these insights indicate that it is possible to influence thermal comfort and temperature perception even in the absence of thermal energy.

Previous work took a first step towards investigating how thermal comfort and thermophysiological responses can be influenced through interactive systems~\cite{KocurHenze23}. The authors immersed users in VR and systematically varied the virtual environment and the embodied avatar. Results suggest that being in a fire world or having fire hands increases the perceived temperature. It was even found that having fire hands decreases the hand temperature compared to having ice hands. Thereby, findings show that thermal comfort and thermophysiological responses can be influenced by interactive systems.

In recent years, mixed reality (MR)---a technology combining elements from VR and augmented reality (AR)---is increasingly adopted in many areas ranging from healthcare, gaming and entertainment, education and training, as well as various research disciplines~\cite{10.1007/978-3-030-05767-1_7}. Considering the ongoing commercial development towards MR, including Microsoft's HoloLens, Meta's Metaverse, and Apple's Vision Pro, experts predict that users will spend considerable amounts of time in MR in the near future~\cite{Park2018}. The use of interactive systems consumes energy and therefore contributes to climate change. Carefully designing immersive experiences to maintain thermal comfort could, however, offset the energy required for cooling or heating the user's physical environment.

With this paper, we propose to systematically investigate how to design immersive experiences to maintain thermal comfort and thereby reduce the energy needed for heating and cooling. Ultimately, we aim to foster HCI's contribution to fighting climate change. Previous work showed what we visually perceive has an effect on our thermal comfort and even our physiological response. Thereby our visual experience influences what we perceive as thermally comfortable. MR and especially VR can visually immerse users in any imaginable environment while embodying any virtual body. MR is expected to be used by a very large number of users on a daily basis. Hence, the overall objective of the proposed experiments is to understand how the design of immersive experiences can influence users' thermal perception and regulation. 

\begin{figure*}
    \centering
    \includegraphics[width=0.49\textwidth]{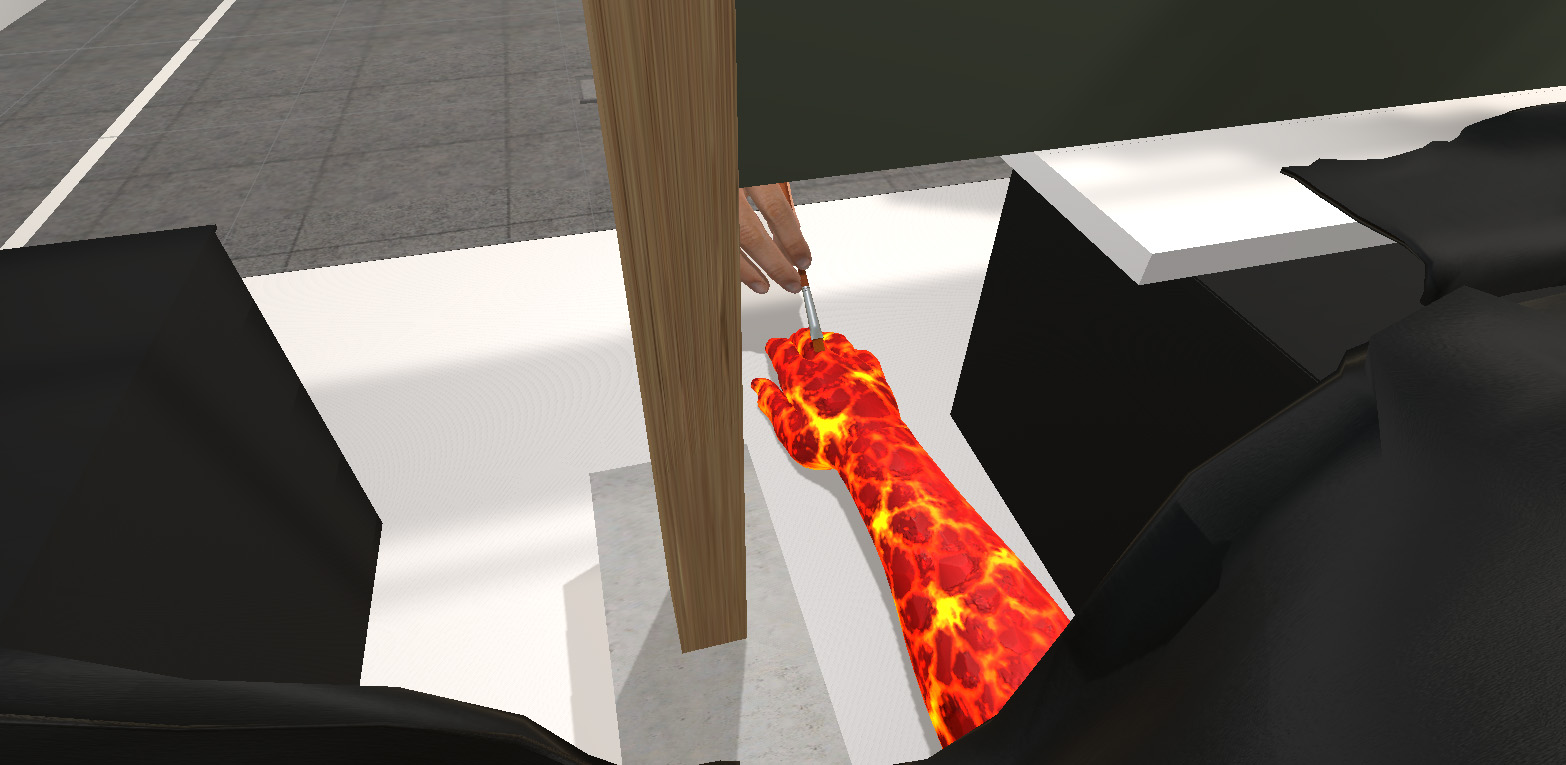}
    \includegraphics[width=0.49\textwidth]{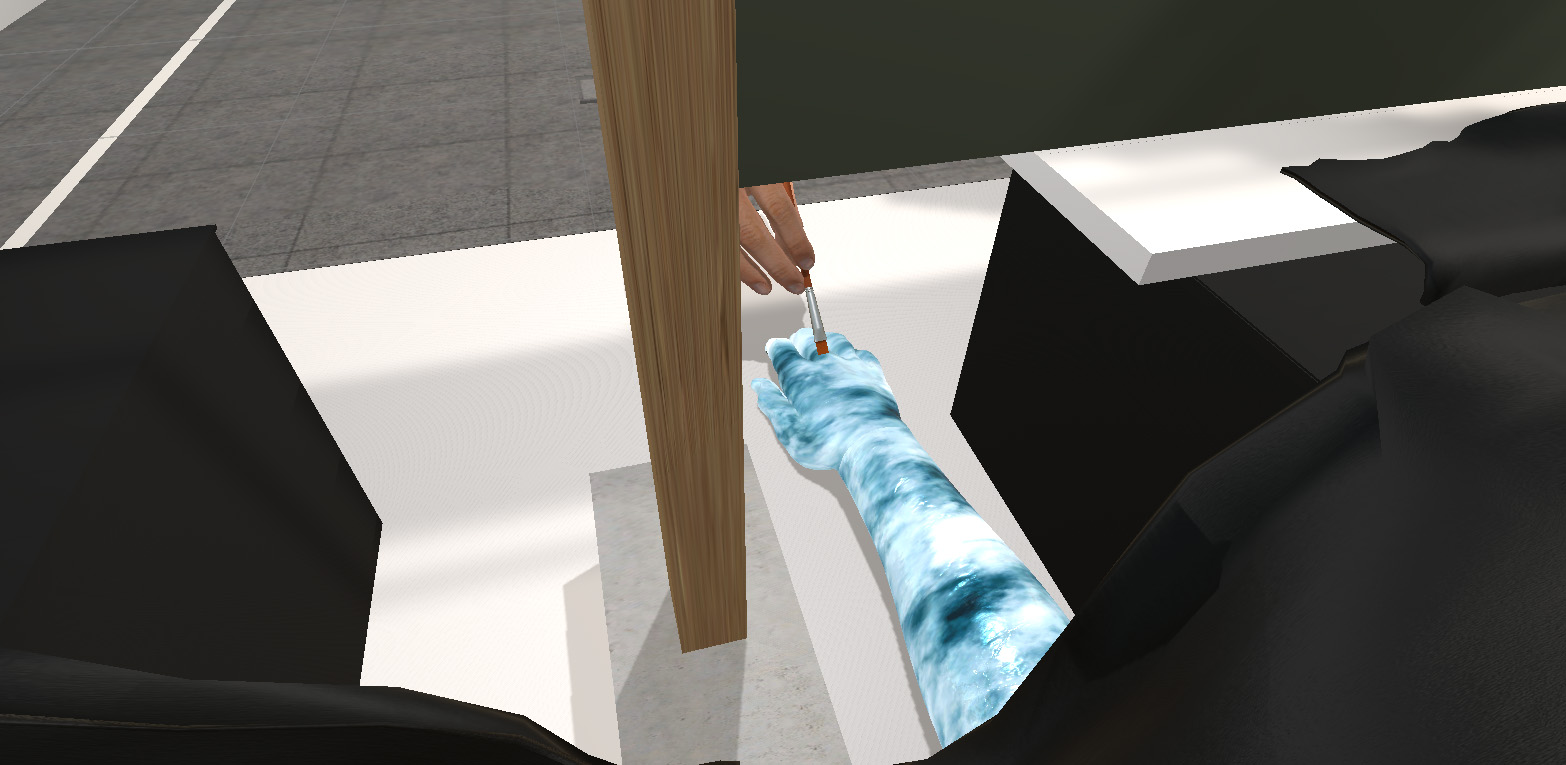}
  \caption{We aim to conduct a rubber-hand illusion experiment~\cite{Botvinick1998,Kocur22RHI} to examine whether the thermal effects induced by the avatar arise from the sense of embodiment in an avatar representing extreme thermal conditions (hot on the left, cold on the right) or simply from the presence of thermal cues, regardless of embodiment sensations. We hypothesize that synchronous stroking of the virtual and real hand will produce greater thermal effects than asynchronous stroking.}
  \label{fig:rhi}
\end{figure*}

\section{Planned Experiments}
In this paper, we propose three experiments that will be conducted in future to learn how immersive experiences can be leveraged to systematically influence thermal comfort and thermophysiological responses. This should, in turn, contribute to a more efficient energy consumption and ultimately support to combat climate change.

\subsection{Experiment 1: Effects of Virtual Heating on Thermal Perception and Skin Temperature}

Previous work found that being immersed in a fire or ice world influences thermal perception (see Fig. \ref{fig:teaser}). However, it is unknown if localized heating or cooling, for example, using virtual heaters also influence thermal reactions. While previous research has shown that thermal cues can influence perception~\cite{KocurHenze23}, the specific effects of virtual heaters on skin temperature and thermal sensation remain unclear. Findings could inform the design of virtual experiences, for example, by leveraging virtual heaters to make participants feel warmer and offset the energy required for heating the surrounding environment.

In this VR experiment, we aim to systematically investigate the effects of virtual heaters on the left and right hand. Understanding whether these effects differ between hands allows to gain insights into localized thermal responses, which may inform future VR interface designs. We plan to employ a repeated-measures design with one independent variable \textsc{Heater} with the two levels \textit{On} vs. \textit{Off}. In each condition, one heater will be activated while the other remains off. We will continuously measure skin temperature for both hands. In addition, we will administer questionnaires in VR that will be completed using gaze to prevent users' hands from being moved and active. We will survey participants for their thermal perception and thermal comfort for the left and right hand individually. To control for potential moderators, we will also assess participants' sense of presence. Fig. \ref{fig:heater} shows the experimental apparatus.

\subsection{Experiment 2: Effects of Embodying Virtual Fire or Ice Hands On Thermal Responses}

Recent work revealed that fire hands increased and ice hands decreased skin temperature in VR~\cite{KocurHenze23}. In line with the Proteus effect~\cite{epub52677,Kocur2020Flex,Sweat,Kalus2023PumpVR,kocur2025investigating}, the visual appearance of the avatar influences users' perception and induces thermophysiological changes. However, it is unkown if the effects are caused and modulated by the extent of experienced embodiment over the hands (i.e., changes in self-perception result in different thermophysiological responses) or the mere presence of warm and cold stimuli (i.e., priming effects)~\cite{Yee2009}. To understand the effects, we aim to harness the rubber-hand illusion (RHI) experiment to systematically modulate the experienced embodiment~\cite{RHIDemo,Kocur22RHI,Botvinick1998}.

In 1998, \citet{Botvinick1998} demonstrated that humans
can experience an artificial limb—a rubber hand—as if it was their own hand. The RHI is a illusory sensation of embodying the rubber hand and induced by multisensory conflicts. When the subjects’ real hand, which is hidden from view, and the rubber hand are stroked at the same time while the subjects can see the rubber hand, after some time they start to perceive the artifical limb as being a part of their own body. As a result, subjects perceive the own hand to be closer to the rubber hand. This phenomenon is known as proprioceptive drift.

Instead of using a rubber hand, we will use a virtual fire and ice avatar hand that will be stroked either synchronously (simultaneously seeing the virtual hand being stroked and feeling the real hand being stroked) or asynchronously (seeing the virtual hand being stroked while feeling the stroke on the real hand with a delay). Hence, we will employ a repeated-measures design with two independent variables \textsc{Hands} (\textit{fire} vs. \textit{ice}) and \textsc{Synchrony} (\textit{synchronous} vs. \textit{asynchronous}). In the \textit{synchronous} conditions, participants should experience a higher sense of embodiment than in the \textit{asynchronous} condition. If the effects are caused by the embodiment, we should therefore evidence larger effects in the \textit{synchronous} condition. We will assess skin temperature, thermal perception, the proprioceptive drift as well as the experienced embodiment. Fig. \ref{fig:rhi} shows the experimental apparatus from participants' point of view.

\begin{figure*}
    \centering
    \includegraphics[width=.42\linewidth,trim = 0cm 0cm 0cm 0cm,clip]{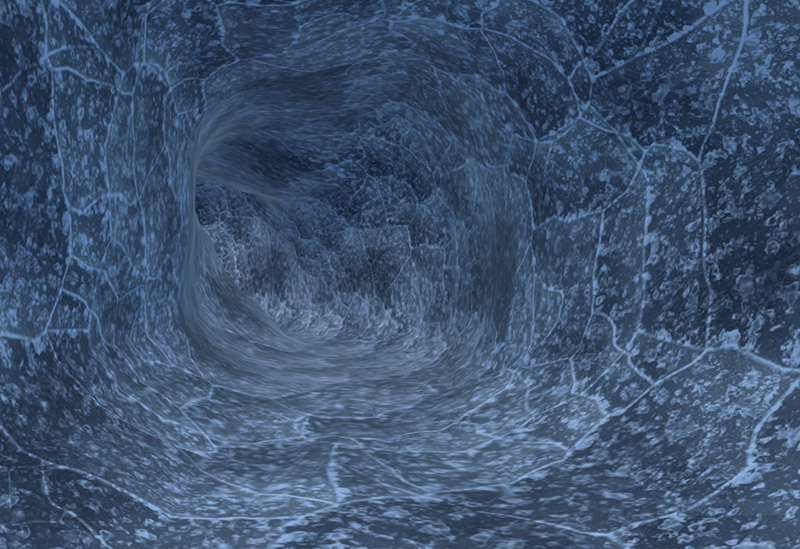}
    \includegraphics[width=.42\linewidth,trim = 0cm 0cm 0cm 0cm,clip]{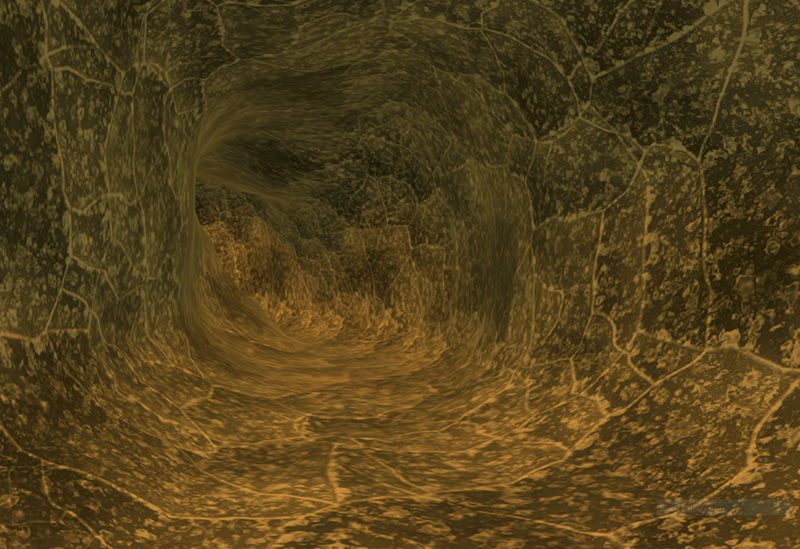}
    \includegraphics[width=.42\linewidth,trim = 0cm 0cm 0cm 0cm,clip]{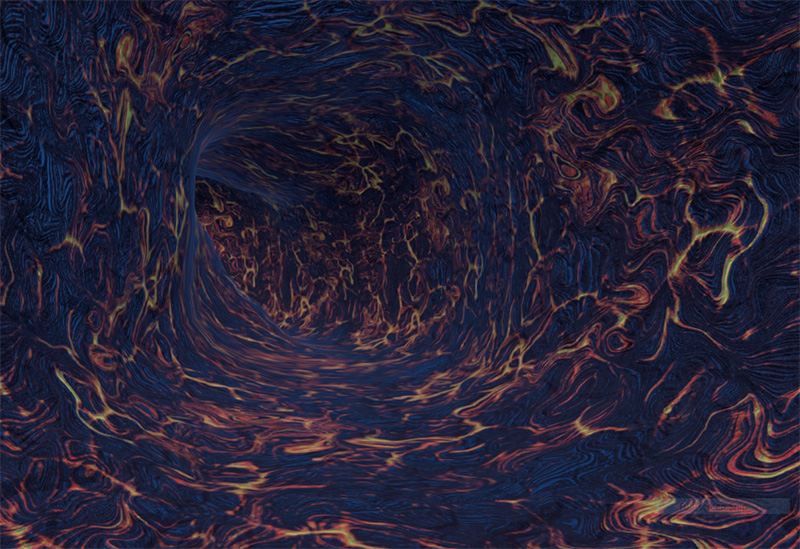}
    \includegraphics[width=.42\linewidth,trim = 0cm 0cm 0cm 0cm,clip]{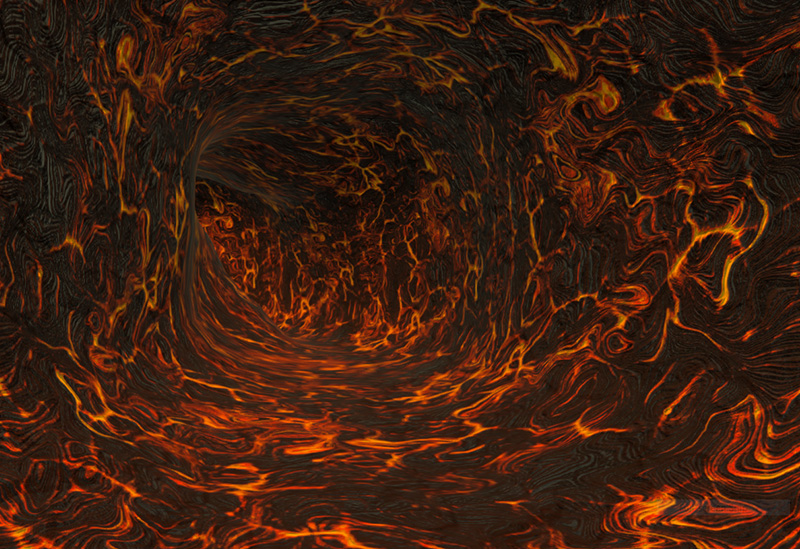}
    \caption{The virtual cave environments with the cold cues (top) and hot cues (bottom) in 12500 K (left) and 4500 K (right) conditions. In this experiment, we plan to explore if thermal effects caused by the virtual environment are induced due to the color temperature or the thermal associations connected with the virtual environment (i.e., fire environment triggers heat associations that in turn changes thermal responses).}
    \label{fig:arexperiences}
\end{figure*}

\subsection{Experiment 3: Disentangling Thermal Effects of Color Temperature and Thermal Cues}
VR enables to immerse users in any possible environment. Consequently, previous work investigated whether virtual environments can also influence thermal perception. Multiple studies found effects of color temperature on perceived temperature~\cite{chinazzo_effect_2017, huang_effects_2019, chinazzo_temperaturecolor_2021, vittori_subjective_2021}. Kocur et al.~\cite{KocurHenze23} recently found that being in a fire world or having fire hands increased the perceived temperature and even showed that having fire hands decreases the hands' skin temperature compared to having ice hands. This is in line with Takakura et al.~\cite{takakura_visual_2013, takakura_nonthermal_2015} who found that watching videos showing hot and cold environments can affect skin and core body temperature. However, neither Kocur et al.~\cite{KocurHenze23} nor Takakura et al.~\cite{takakura_visual_2013, takakura_nonthermal_2015} controlled for the hue of the displayed scene. Scenes with hot or cold visual thermal cues were presented but the scenes also differed in their color temperature. Thus, it is unclear if the effects were caused by the thermal cues or just by the color temperature of the displayed scene.

In this experiment, we, therefore, plan to disentangle the effects of thermal cues and the color temperature by controlling for both. Fig. \ref{fig:arexperiences} shows the cold and hot thermal cues (ice environment and fire environment) in a cold hue (12500 K) and a hot hue (4500 K). We will employ a repeated-measures design with two independent variables \textsc{thermal cue} (\textit{hot} vs. \textit{cold}) and \textsc{Hue} (\textit{4500~K} vs. \textit{12500~K}). Participants will be immersed in the respective condition while continuously measuring skin temperature. After each condition, participants will be asked to assess their thermal perception and comfort in VR.

\section{Summary}
In this paper, we propose three experiments to explore how the design of immersive experiences can induce thermal effects to mitigate increased energy consumption or even encourage users to reduce their own energy use. As immersive technology becomes increasingly prevalent in daily life, leveraging thermal cues to enhance thermal comfort—without relying on actual thermal energy—presents a promising opportunity. Understanding these effects is crucial for optimizing virtual experiences and promoting sustainable energy practices.

Experiment 1 investigates localized heating through virtual heat sources. Experiment 2 builds on previous research by examining whether the degree of virtual embodiment of avatar hands modulates thermal effects. Experiment 3 aims to determine whether thermal effects in virtual environments stem from color temperature itself or from cognitive associations with thermal cues.

Together, these three experiments pave the way for future research by deepening our understanding of virtual thermal perception. This knowledge can have significant implications not only for improving VR immersion but also for addressing climate change by informing energy-efficient design strategies and promoting behavioral shifts toward reduced energy consumption.
\bibliographystyle{ACM-Reference-Format}
\bibliography{sample-base}

\end{document}